\documentclass[twocolumn,showpacs,prl]{revtex4}
\usepackage[pdftex]{graphicx}
\usepackage{epstopdf}
\usepackage{dcolumn}
\usepackage{bm}

\def\gappeq{\mathrel{ \rlap{\raise.5ex\hbox{$>$}}
                      {\lower.5ex\hbox{$\sim$}}  } }
\def\lappeq{\mathrel{ \rlap{\raise.5ex\hbox{$<$}}
                      {\lower.5ex\hbox{$\sim$}}  } }

\begin{document}

\preprint{PRA}

\title{Comparison of Hard-Core and Soft-Core Potentials for Modelling Flocking in Free Space}

\author{J.A. Smith and A.M. Martin.}

\address{School of Physics, University of Melbourne, Parkville, Victoria 3010, Australia.}

\date{\today}

\begin{abstract}
An investigation into the properties of a two dimensional (2D+1) system of self propelled particles (known as boids) in free space is conducted
using a Lagrangian Individual-Based Model.  A potential, associated with each boid is specified and a Lagrangian is subsequently derived in order to obtain the equations of motion for each particle in the flock.  The Morse potential and the Lennard-Jones potential, both well understood in atomic and molecular physics, are specified.  In contrast to the original model proposed by Vicsek [Phys. Rev. Lett.
{\bf 75}, 1226 (1995)] systems are considered with open boundary conditions.  These two models successfully replicate the phases observed in Vicsek's original model, as well as several other significant phases, providing a realistic model of a wide range of flocking phenomena.
\end{abstract}

\pacs{05.65.+b, 87.10.-e, 87.18.Vf}
\maketitle

\section{Introduction}

The concept of spontaneous flocking is familiar to us all \cite{Parrish99,NatVic,Erdmann08}.  The world around us
is teeming with examples of this strange emergent behaviour, from a
flock of birds to a school of fish \cite{Huth}, a herd of wildebeest\cite{Couzin02} to a swarm
of locusts \cite{Buhl06,Buhl08}. In Physics, a flock is defined
as the coherent motion of a group of self-propelled particles emerging from a simple set of interactions between the constituents of that group.
Tam\'as Vicsek first introduced the concept of flocking to the physics
community in 1995 \cite{Vic}, through the consideration of a simple agent-based model, paving the way for a new
branch of interdisciplinary research into the physical mechanisms
which result in the emergence of flocking \cite{Czi,Csahok95,Toner95,Hemmingsson95,Albano96,Czirok97,Czirok99,Czrok99a,Mikhailov99,Gregoire01,Gregoire03,Huepe04,Ton,Tu,Chate06,Aldana07,Chate08}.
\\

In his paper, Vicsek and others discovered that a group of self propelled particles, when modelled using a simple set of parameters displayed the same characteristics as a ferromagnetic system.  At high values for stochastic noise, the motion of the particles (known in flocking research as boids) was essentially random, analogous to a high temperature ferromagnetic system.  As stochastic noise is reduced past a critical value, a phase transition occurs, whereby the direction of the boids suddenly becomes correlated, and their direction of motion aligns.  This simple observation gave birth to the field of flocking.  Since this time physicists have gained a deeper understanding of the analogies and relationships between physical condensed matter systems, and biological self-propelled systems.\\

A number of key states have been observed and analyzed in flocking research since the original discovery by Vicsek \emph{et al}.  They include moving and stationary fluids, moving and stationary solid states, and single vs multiple flock states \cite{Buhl06,Vic,Czi,Czirok97,Gregoire03,Huepe04,Ton,Tu}.
The phase transitions which have received the most attention are the breaking of directional symmetry \cite{Buhl06,Vic,Czi,Czirok97,Gregoire03,Ton,Tu}, and the breakdown from a single flock into multiple independent flocks \cite{Huepe04,Ton,Tu}.  The evolution of vortices has also recieved significant attention, particularly in recent years \cite{Hemmingsson95,Orsogna06,damnyou,Bob}.
The formation of vortices, rigid rotational states have also been observed \cite{Bob} and can be physically related to the behavior of some bacteria such as \emph{Dictyostelium} and \emph{Daphnia} which have been observed experimentally to form vortices \cite{Rappel,Daph,Ebac}.\\

One of the major short-comings of the approaches adopted thus far
is that they have relied on the specification of periodic \cite{Vic,Czi,Csahok95,Toner95,Czirok97,Czirok99,Czrok99a,Mikhailov99,Ton,Tu,Gregoire01,Gregoire03} or reflecting \cite{Hemmingsson95,Albano96}
boundary conditions to hold the flock together.  The
surface of the flock is imposed artificially rather than being allowed to arise
naturally out of interactions between the boids themselves. This has
meant that although many of the flocks bulk properties are well understood,
the surface properties of the flock have yet to be studied in detail \cite{Ton}.\\

In this paper, two non-local flocking models, based on the Morse and Lennard-Jones potentials, are investigated in free space.  Whilst some research has now begun on flocking in free space \cite{Orsogna06,damnyou,freespace}, most flocking models continue to be based on periodic boundary conditions, which places restrictions upon their applicability to real world situations.  Additionally, many of the newer models in free space focus on the specific modelling of bacterial systems, in particular, the simulation of vortex states which have been observed in a number of bacterial colonies, as mentioned above\cite{Orsogna06,damnyou,Bob,Rappel,Daph,Ebac}.  The model developed here is designed to be a simple, robust and independent of scale. This makes it capable of accurately simulating a large number of biological systems, over a wide range of length scales, from bacterial colonies right up to ungulate herds.  Rather than attempting to model a specific instance of flocking behaviour very accurately, the models proposed will instead be aimed at exploring the generality of flocking behaviour, clearly showing how many of the characteristics of seemingly quite different grouping behaviours, from the very small to the very large scale, can in fact be encapsulated through a single model with a small number of free parameters.  Thus flocking can be seen as a universal behaviour with a unique set of characteristics which transcend the particular biological organisms which are displaying the behaviour. \\

A Lagrangian approach is taken in order to derive the equations of motion, and the flock is then analyzed as a non-local field, where each boid makes a contribution to the potential of the region. A comparison is made between a hard-core Lennard-Jones Potential and a soft-core Morse potential to determine the similarities and differences observed when modelling flocking using these two potentials.  The Lennard-Jones and Morse potentials provide suitable features for modelling flocking and are well understood in the context of atomic and molecular physics.\\

\section{Model Formulation}

Tam\'as Vicsek introduced flocking to the physics community in his
paper entitled {\it Novel Type of Phase-Transition in a System of
Self-Driven Particles } \cite{Vic} where he showed the emergence of
ordered states from a disordered initial condition.  Starting with a 2D box bounded periodically, particles were distributed at random positions
${\bf r}_i$ with a constant speed $|{\bf v}_i|$ in some direction
$\theta_i$.  At each time step, separated by $\Delta t$, the new position and direction, of each boid,  were
given by:
\begin{equation}
\textbf{r}_i(t+\Delta t)= \textbf{r}_i(t)+\textbf{v}_i\Delta t,
\label{eq:a1}
\end{equation}
\begin{equation}
\theta_{i}(t+\Delta t)= \langle \theta_i(t) \rangle_r + \Delta \theta,
\label{eq:a2}
\end{equation}
where $\langle\theta_i(t) \rangle_r$ defines the average angle of
all the boids within  a distance $r$ of boid $i$ and $\Delta \theta$ corresponds to a random number chosen from a uniform distribution between $-\eta/2$ and $\eta/2$.  Despite the importance of this model to flocking research in physics, it does have a number of short-comings when used to model realistic flocking systems.  Two key issues are the lack of short-range repulsion to prevent particle collision and the reliance on periodic boundary conditions to effectively force continued interaction over long time periods and prevent dispersion.  This means that the full range of realistic flocking behavior is not observed using this model.\\

A number of different approaches have been attempted in order to overcome these shortcomings.  Often, the approach taken includes the ad-hoc introduction of some short range repulsion or alignment dependent equations of motion \cite{Vic,Csahok95,Hemmingsson95,Albano96,Czirok97,Czirok99,Czrok99a,Gregoire01,Gregoire03}.  The approach taken here is similar, but instead follows in the footsteps of \cite{Huth,Tu,Orsogna06,damnyou,Lee} by introducing an already well understood potential, and then deriving the associated equations of motion.  In this instance, two different non-local potentials are chosen, in order to contrast the behaviour of flocks based on hard-core and soft-core potentials. The two potentials which are used are the Lennard-Jones Potential (LJP) and the Morse Potential (MP) which are commonly used molecular potentials.  After a randomized initial placement these systems are allowed to evolve in free space, allowing a wider range of possible flocking phenomena to occur.  This approach extends previous studies \cite{Huth,Tu,damnyou,Orsogna06}, by consideration of open boundary conditions and direct comparison of the effects of hard (LJP) and soft (MP) core potentials on such parameters as flock density, cohesion and behaviour around the critical points.\\

\begin{figure}
\centering
\includegraphics[width=8.5cm]{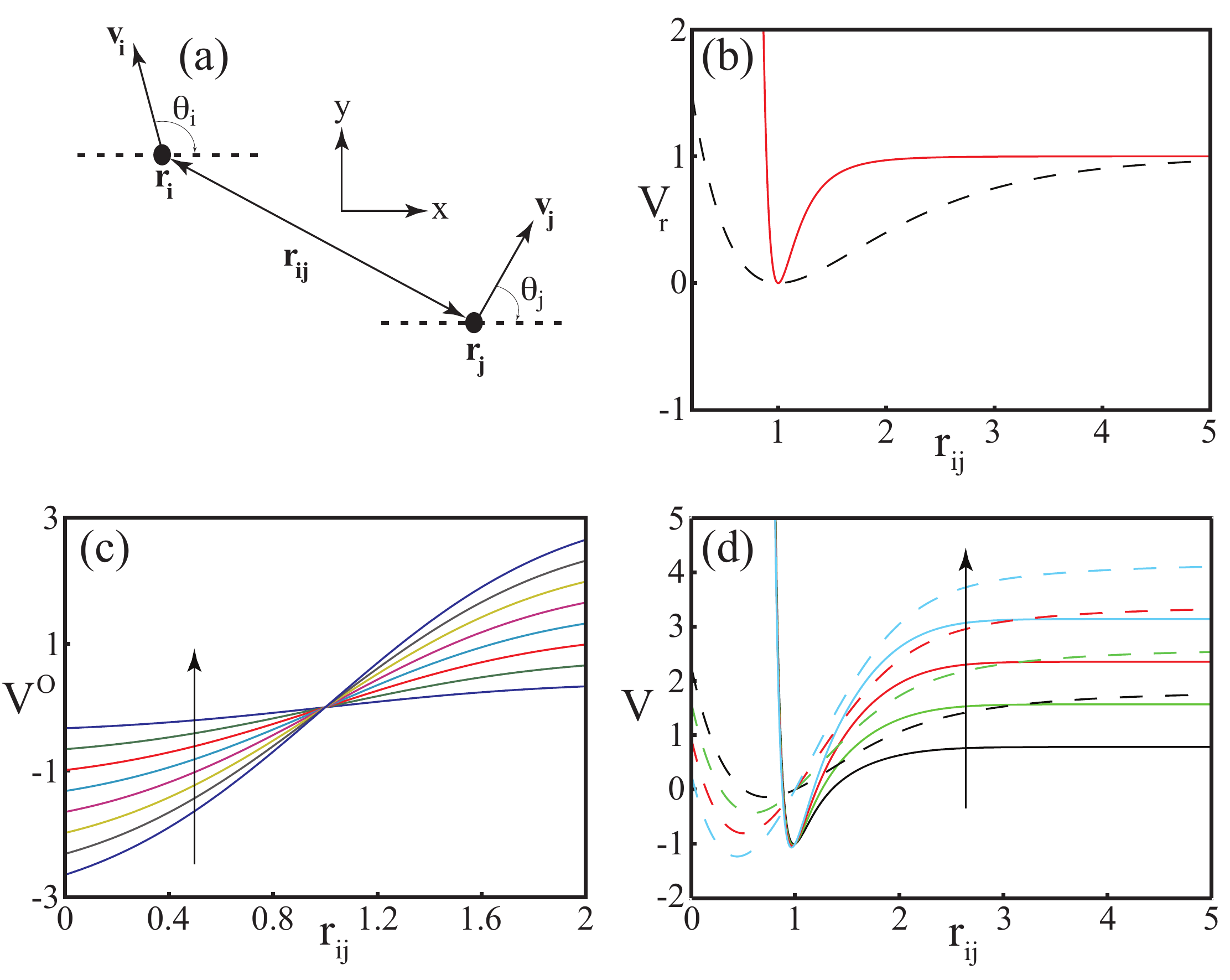}
\caption{(a) Specification of parameters:  a representation of two boids positioned at ${\bf r}_i$ and ${\bf r}_j$ with respective velocities ${\bf v}_i$ and ${\bf v}_j$ in directions $\theta_i$ and $\theta_j$, where ${\bf r}_{ij}= {\bf r}_i - {\bf r}_j$. (b) Lennard-Jones (red solid line) and Morse (blue dashed line) Potentials. (c) Orientation Potential for values of $\theta$ increasing in the direction of the arrow, between $-\pi$ and $\pi$ in increments of $\pi/8$, where $C = \alpha = \beta = 1$. (d) Total potential $V = V_r + V^{{\rm O}}$ for values of theta between $-\pi$ and $\pi$ in increments of $\pi/4$ (increasing in the direction of the arrow), for the Lennard-Jones (solid lines) and Morse (dashed lines) Potentials. \label{Fig:Potentials}}
\end{figure}

First, the interactions between two boids, $i$ and $j$  are considered, as shown in Fig. \ref{Fig:Potentials}(a). $r_{ij}$ is the distance
separating any 2 boids while ${\bf v}_i$ and ${\bf v}_j$ are the velocities of boids $i$ and $j$, with their direction of motion being defined as  $\theta_i$ and $\theta_j$, as measured from the horizontal.\\

To fully encompass the essential properties of flocking three key elements are required.  These can be encapsulated through the consideration of two boids separated by distance $r_{ij}$.  Boids have a tendency to avoid collisions thus a repulsive force is required for $r_{ij} < r_{{\rm min}}$.  Boids have some preferred separation $r_{ij} = r_{{\rm min}}$, a balance between collision avoidance and predation risk. Boids cease to interact for large separations.  Both the LJP and MP, widely used for modelling in condensed matter physics, satisfy the above criteria as can be seen in Fig. \ref{Fig:Potentials}(b). These potentials are qualitatively  similar for $r_{ij} > r_{{\rm min}}$ but possess a significant quantitative difference for $r_{ij} < r_{{\rm min}}$.  Specifically, as $r_{ij}\rightarrow 0$ the LJP has a hard-core with $V\rightarrow \infty$ whereas  the MP has a soft-core with the potential tending to a finite constant.  This difference generates different bulk behaviour in the flock and also effects the behaviour of the flock around the critical points.\\

The LJP is given by:
\begin{equation}
V_{i}^{{\rm LJ}}({\bf r})=
\sum_{j \ne i}\epsilon\left[\left(\frac{r_{{\rm min}}}{
r_{ij}}\right)^{12}-2\left(\frac{r_{{\rm min}}}{
r_{ij}}\right)^6\right], \label{eq:b1}
\end{equation}
where $V_{i}^{{\rm LJ}}$ is the potential between boid $i$ and the rest of the flock,
${r}_{ij}$ is the distance between boid $i$ and boid $j$ in the flock.
$r_{{\rm min}}$ and $\epsilon$ are constants associated with the shape of the potential.
The MP is specified as:
\begin{equation}
V_{i}^{{\rm M}}({\bf r})=
\sum_{j \ne i}\epsilon[1-\exp{(-ar_{ij}-r_{{\rm min}})}]^2, \label{eq:b1}
\end{equation}
where $a$ is a constant defining the width of the potential well.  In each case the depth of the potential is defined by $\epsilon$ and the minimum potential between the two boids occurs at $r_{{\rm min}}$, which may be physically interpreted as the preferred distance between nearest neighbours in the flock.  The functional form of the LJP (red solid line) and MP (blue dashed line) are shown in Fig. \ref{Fig:Potentials}(b) for the case of two boids.\\

The final key element for flocking behaviour is alignment. For two
boids $i$ and $j$ to be aligned, the relative angle between them
must be $\Delta\theta_{ij}=\theta_{i} - \theta_{j} \approx0$. In order to enforce this kind of
alignment an angular dependent potential, $V_i^{{\rm O}}(\textbf{r})$ is introduced, which we
henceforth call the 'orientation potential'. This potential is fairly localized, as the individual
boids only interact with nearest  neighbours and are unable to
discern the flock as a whole.  This means that any boid will attempt
to match the direction of it's nearest neighbours, but boids far
away will have little effect.  A convenient representation of such a potential is the error function:
\begin{equation}
V_{i}^{{\rm O}}({\textbf{r}}) = \sum_{j \ne i}C\theta_{j}\times
{\textrm{erf}}{{((\alpha r_{{\rm min}}-r_{ij})/\beta)}},\label{eq:Orient}
\end{equation}
where C, $\alpha$ and $\beta$ are constants determining magnitude of the potential, the position of the minima and the width of the potential respectively.  In Fig. 1(c) the functional form the orientation potential is shown for two boids as a function of the orientation angle $\theta$, between them.  $\alpha$ and $\beta$ from equations (\ref{eq:c11}) and (\ref{eq:c12}) are suitibly chosen to be $2$ and $1$ respectively as this value provides a maximum direction correlation at $2r_{{\rm min}}$ ensuring that the interaction is short range only.\\

$V_{i}^{{\rm LJ}}({\bf r})$ and $V_{i}^{{\rm O}}({\bf r})$ are added to obtain a potential of the form:
\begin{eqnarray}
V_i({\bf r}) &=& \sum_{j \neq i}\epsilon\left[\left(\frac{r_{{\rm min}}}{
r_{ij}}\right)^{12}-2\left(\frac{r_{{\rm min}}}{
r_{ij}}\right)^6\right] \nonumber \\
&+&
C\beta \theta_{j}\times
{\textrm{erf}}\left(({\alpha r_{{\rm min}}-r_{ij})/\beta}\right).
\label{eq:b2}
\end{eqnarray}
and likewise for $V_{i}^{{\rm M}}({\bf r})$ and $V_{i}^{{\rm O}}({\bf r})$:
\begin{eqnarray}
V_i({\bf r}) &=& \sum_{j \neq i}\epsilon[1-\exp{(-ar_j-r_{{\rm min}})}]^2 \nonumber \\
&+&
C\beta \theta_{j}\times
{\textrm{erf}}\left(({\alpha r_{{\rm min}}-r_{ij})/\beta}\right).
\label{eq:b3}
\end{eqnarray}
The functional form of $V_i(r)$ for 2 boids $i$ and $j$ are shown in Fig. 1(d).  The different natures of the two potentials at small values of $r_{ij}$ can also be clearly observed Fig. 1(d), the MP goes to a finite value as $r\rightarrow 0$ whereas the LJP extends to infinity as $r\rightarrow 0$.\\

The force on each boid is derived using a Lagrangian approach:
\begin{eqnarray}
L= T - V,
\end{eqnarray}
where $L$ is the Lagrangian of the system, $T=\frac{1}{2}m\dot{{\bf x}}^2 + \frac{1}{2}m\dot{{\bf y}}^2$ is the kinetic energy of the system, and $V$ are the potentials described above.  Using the generalized Euler-Lagrange equations in $x$ and $y$:
\begin{eqnarray}
\frac{d}{dt}\left(\frac{dL}{d\dot{x}_i}\right) - \frac{dL}{dx_i} +
\frac{d\Lambda_{x_i}}{d\dot{x}_i}=0,\label{eq:c2}
\end{eqnarray}
\begin{eqnarray}
\frac{d}{dt}\left(\frac{dL}{d\dot{y_i}}\right) - \frac{dL}{dy_i} +
\frac{d\Lambda_{y_i}}{d\dot{y}_i}=0,\label{eq:c2}
\end{eqnarray}
\\
with $\Lambda_{i} = \frac{1}{2}\lambda\dot{{\bf x}}_i^2 - \frac{1}{4}\omega \dot{{\bf x}}_i^4 + \frac{1}{2}\lambda\dot{{\bf y}}_i^2 - \frac{1}{4}\omega \dot{{\bf y}}_i^4$.\label{eq:c4}
$\Lambda_{x_i}$ and $\Lambda_{y_i}$ incorporate both a self damping and a self acceleration term, which effectively cause the flock to approach an average speed of $\upsilon = \sqrt{\frac{\omega}{\lambda}}$ \cite{Mikhailov99,damnyou} whilst still allowing variation about this speed from individual boids within the flock.  Introducing damping in this way allows for a great deal of control over the flocks velocity, increasing the stability and cohesion of the flock.\\

From the Lagrange Equations, the equations of motion for the LJP are:
\begin{eqnarray}
{\bf F}_i({\bf {r}}) = -\sum_{j\neq i}{12\epsilon\over{r_{{\rm min}}}}\left(\left(\frac{r_{\rm min}}{
r_{ij}}\right)^{13}-\left(\frac{r_{{\rm min}}}{
r_{ij}}\right)^7\right)\hat{r}_{ij} + \nonumber \\
A{\bf \theta}_j \exp{\left(\frac{-(r_{ij}-\alpha
r_{min})^2}{\beta^2}\right)}\hat{r}_{ij}-(\lambda-\omega v^2)\dot{{\bf r}_i}+\eta_i,\label{eq:c11}
\end{eqnarray}
and likewise for the MP:
\begin{eqnarray}
{\bf F}_i({\bf {r}}) = \sum_{j \neq i}2a\epsilon[1-e^{{(-ar_j-r_{{\rm min}})}}]e^{{(-ar_j-r_{{\rm min}})}}\hat{r}_{ij} + \nonumber \\
A{\bf \theta}_j \exp{\left(\frac{-(r_{ij}-\alpha
r_{{\rm min}})^2}{\beta^2}\right)}-(\lambda-\omega v^2)\dot{{\bf r}_i}+\eta_i.\label{eq:c12}
\end{eqnarray}
Note that the new constant $A=\frac{4}{\beta\sqrt{\pi}}$ is defined for convenience and that $m=1$ is set, giving all boids equal weighting.  Note that by setting $m=1$ there is a loss of generality, since no individual bird can be set to have a greater weighting than any other (ruling out the modelling of behaviours where there is individual leadership).  Variation of $\lambda$ and $\omega$ can be used to adjust the average velocity of the flock.  These equations now represent a set of equations of motion which can be used to model flocking in free-space on length scales defined by $r_{\rm min}$.\\

There are $4$ important parameters which can be varied to simulate different flocking phases, $A$, $\epsilon$, $\eta$ and the initial density $\rho$.
However, $\epsilon$ is held constant throughout our models, and $A$ is only varied between $0$ and a finite number, in this case $1$, which equates to turning the alignment component of the potential on a off.  This leaves $\eta$, the stochastic noise and $\rho$ the initial density, which are varied with respect to one another in order to examine the range of flocking behaviours observed within this parameter space.\\

In order to measure different states of the system and determine the locations of any phase transitions a set of 4 order parameters are defined, these being, (i) the average separation, $\bar{x}$, (ii) the fluctuation in the average separation, $\Delta \bar{x}$, (iii) the average velocity, $\bar{{\bf v}}$ and (iv) the fluctuation in the average velocity, $\Delta \bar{{\bf v}}$.\\

\section{Results and Discussion}

Initially, each boid is placed randomly within a box of side length $l$ at position $r_i$.  Each boid starts with some randomized initial speed $v_i$ in some angular direction $\theta$, where $\theta$ is as defined in Fig. $1(a)$.  The initial density is thus simply calculated as,
\begin{equation}
\rho_{init}= \frac{1}{l^2},\label{eq:Density}
\end{equation}

In the case of the LJP, an additional restriction that $r_{{\rm init}} > r_{{\rm min}}$ is imposed such that no boids start within the infinite repulsive region of the LJP.  Earlier experimentation found this restriction to be necessary since any boid beginning the simulation within the repulsive region of the Lennard-Jones potential will have extremely high energy with respect to the other boids \cite{Hons}.  This leads to flock breakdown and dispersion in almost all cases considered.  Note that this is one of the major weaknesses in the LJ description of flocking presented here, and does result in a lose of generality when compared to the MP description.  It is also a particularly significant modification since it is found that the future evolution of the flock is critically dependent on the initial flock density.  Each simulation contains $200$ boids and is performed over a total of $20000$ time steps. At each time step, the new position of boid $i$ is calculated using simple newtonian kinematics.
For each set of suitable parameters an ensemble of $100$ runs was performed, and the average of each of the order parameters
listed in the previous section are taken over the entire ensemble.\\

Key phases are then defined in terms of the order parameters that are tracked in the simulation.  The non-moving state is defined by $\bar{{\bf v}}=0$ while the moving state is defined by $\bar{{\bf v}}\neq0$. To differentiate between liquid and solid, the fluctuation in the average separation of the boids in the flock is analyzed.  If $\Delta \bar{x} \sim 0$ then the state is a solid, if $\Delta \bar{x} > 0$ so long as  $\bar{x}\sim constant$ over time, indicating
that the flock remains together and is not undergoing significant dispersion.\\
\begin{figure}
\begin{center}\scalebox{0.345}{\includegraphics{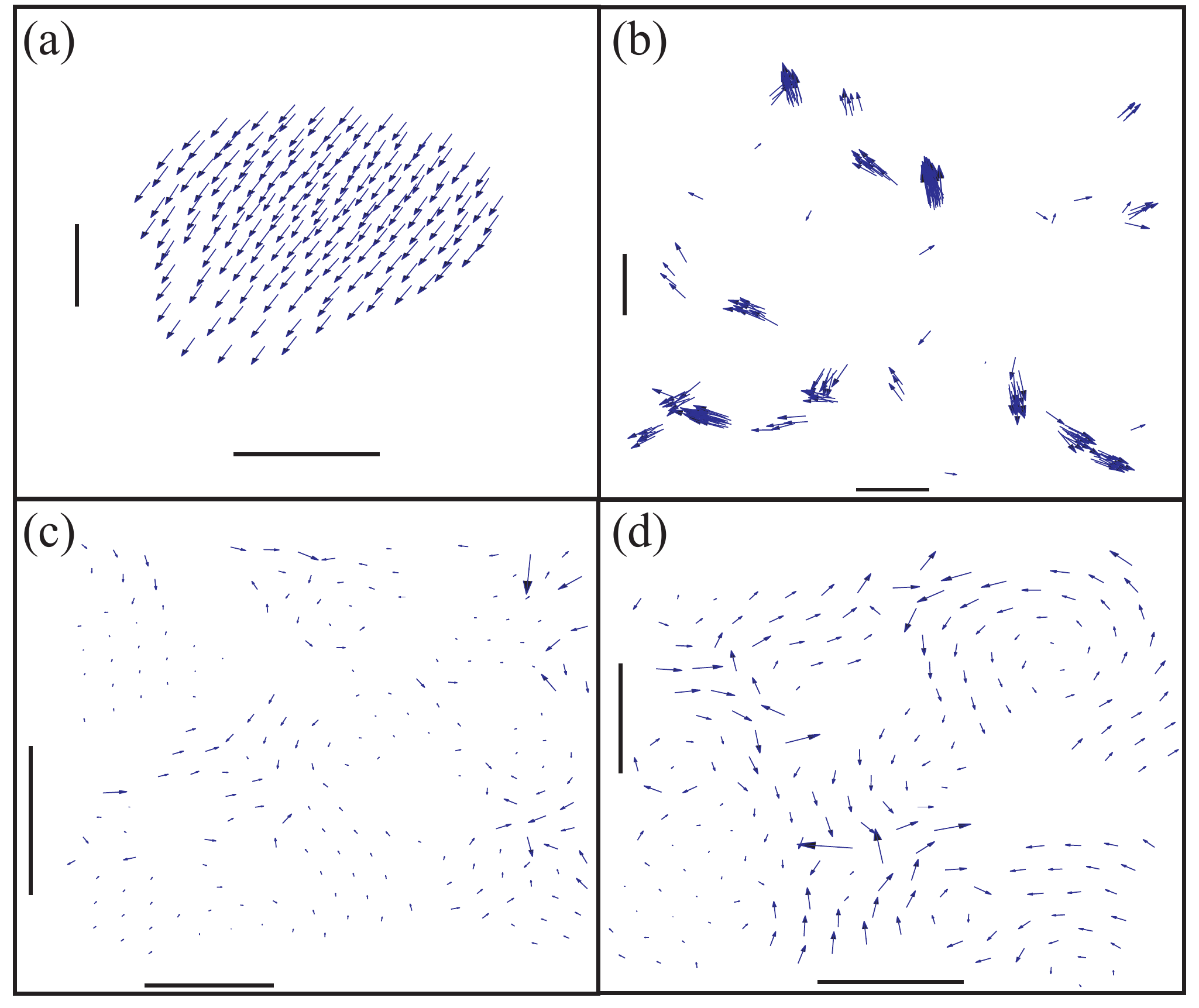}}
\caption{Examples of 4 different coherent states observed for the MP. (a) Single moving flock (b) Multiple Independent moving flocks (c) Stationary lattice-like flock (d) Stationary Liquid flock (with vortex-like behaviour).}
\end{center}
\end{figure}
The phases observed in Vicsek's original model as well as moving and stationary variants of the fluid and solid phases in both the multi-flock and single flock limits were successfully replicated using both soft-core (MP) and hard-core (LJP) potentials in free space by varying the noise amplitude, $\eta$, the initial density, $\rho$ for stationary $(A=0)$ and moving $(A=1)$ flocks.  Importantly, this shows that both models demonstrate similar qualitative behaviour, since neither $\eta$ nor $A$ effect the LJ or Morse components of the potential, only the orientational component.  Since the same kinds of phases are observed in similar relative positions of phase space for both the LJP and MP it indicates a "Potential Independence" between different models.  This suggests that any potential with some short-length repulsion, a single minimum at $r_{min}$ and no long range interactions will likely produce the same qualitative behaviour.\\

Fig. 2 shows a sampling of some of some of the key phases that have been documented in flocking literature and which are observed in this model.  Fig. 2(a) shows a single flock, Fig. 2(b) a phase consisting of multiple independent flocks in a low density regime, Fig. 2(c) a stationary lattice and Fig. 2(d) a stationary liquid exhibiting vortex like structures within the bulk.  It is important to note that all these states were observed in free space, in the absence of any boundary conditions.  Whilst the structures observed in Fig. 2(a) and Fig. 2(b) have been studied in previous models, few of these models have been examined in free space.  Additionally, experimental observation and measurement is significantly lacking for structures like these \cite{Couzin02}, which tend to form on macroscopic levels.  However, the recent advent of the STARFLAG project means that experimental verification for some of these phases may soon be forthcoming \cite{STARBOOK,STARFLAG}.  However the state shown in Fig. 2(d) is in very good qualitative agreement, not just with several previous models looking at complex bacterial behaviour \cite{Bob,Rappel,Daph}, but also with experimental measurements made on E.coli \cite{Ebac}.\\

Despite significant qualitative similarities in structure and behaviour between the two models there are still some key differences between
flocks formed.  Quantitative measurements such as inter-boid distance, position and nature of the phase transitions
and the relative regions of phase space where these different phases are present are still dependent on the finer structure of the different potentials and some variation between MP and LJP cases.  For instance, in the single flock regime, the MP forms a dense, regular teardrop shaped flock, whereas the LJP leads to a flock with a dense frontal arc with a more dispersed region trailing.  Additionally, the positions and nature of the phase transitions between the single flock and multi flock phases is model dependent.  Using the MP, the single-multi transition appears to be discontinuous, effectively occurring along one plane in phase space.  However in the case of the LJP, the single-multi transition occurs gradually, and there is not a sudden change between the single and multi flock phases.  Instead
the flocks begin to disperse at some value of $\rho$ but do not become fully independent until a much higher $\rho$, with a transitionary region existing in between.\\
\begin{figure}
\begin{center}\scalebox{0.345}{\includegraphics{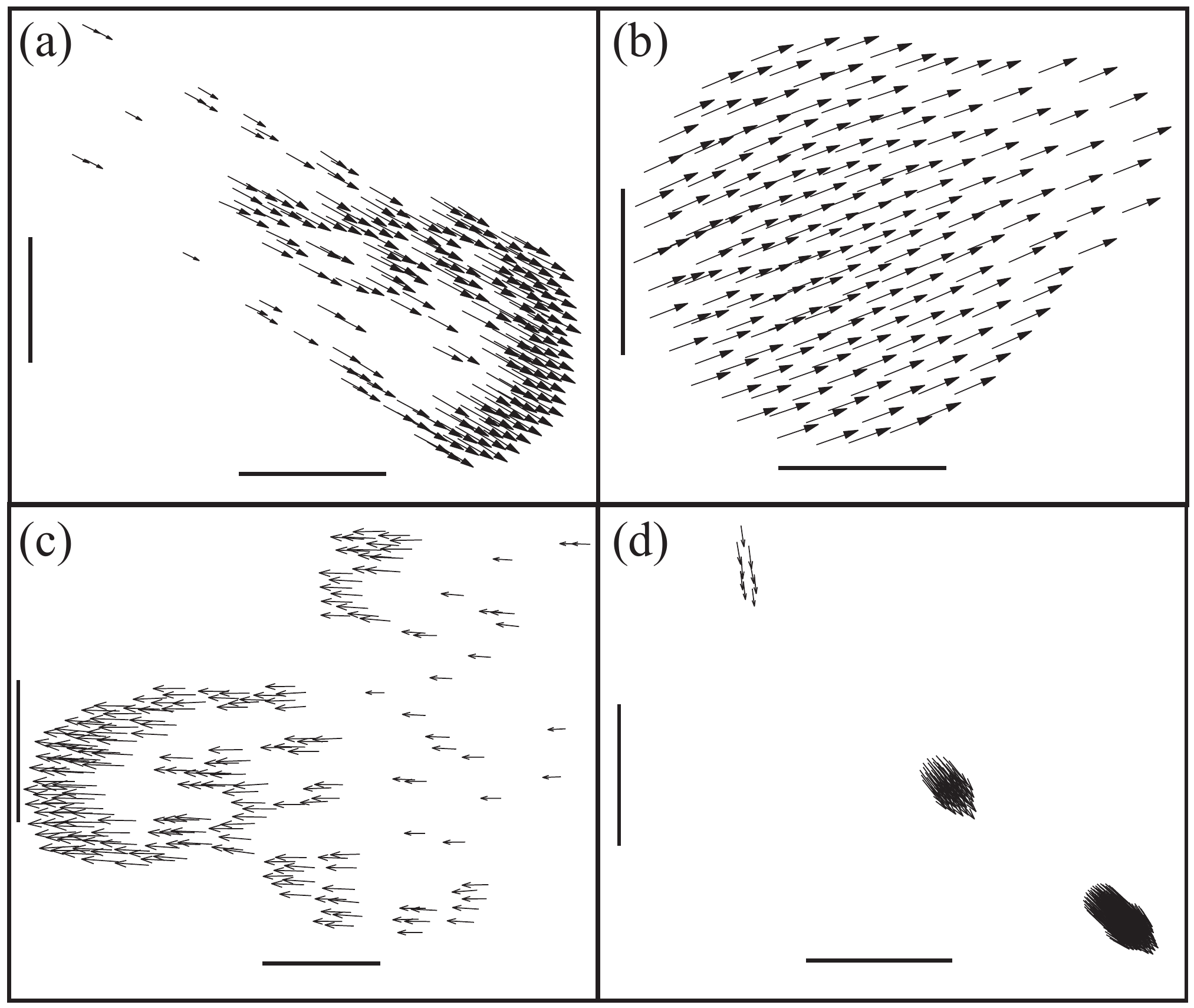}}
\caption{Examples of breakdown modes in MP and LJP flocks. (a) Single moving flock governed by LJP (b) Single moving flock governed by MP (c) Fragmentation of LJP flock demonstrating characteristic "tearing" (d) Fragmentation of MP flock demonstrating the characteristic "train" behaviour.}
\end{center}
\end{figure}
\\
The nature of the transition of the single flock state is highly dependent on the model being used, with a significantly different modes being observed for each model.  In the LJP, the mode by which the single flock transitions into multiple flocks is quite different to that which is observed in the MP case.  The front of the flock begins to split as density decreases, forming two separate "wavefronts" (the term which is adopted to describe this mode hereafter).  The evolution of this fragmentation with decreasing density is seen clearly by comparing figure 3(a) to figure 3(c).  In figure 3(a) there is a single, cohesive front, whereas in figure 3(c) this single front has began to fragment into two or more separate fronts.  After splitting,  each front then continues in a slightly different direction to the other, and over long time scales, the single flocks resolves itself into two separate flocks.  As density decreases further, the two separate flocks become easier to define, until at very low densities, many completely independent flocks can be observed.\\
\begin{figure}
\begin{center}{\includegraphics[height=7.8cm, width=5.7cm]{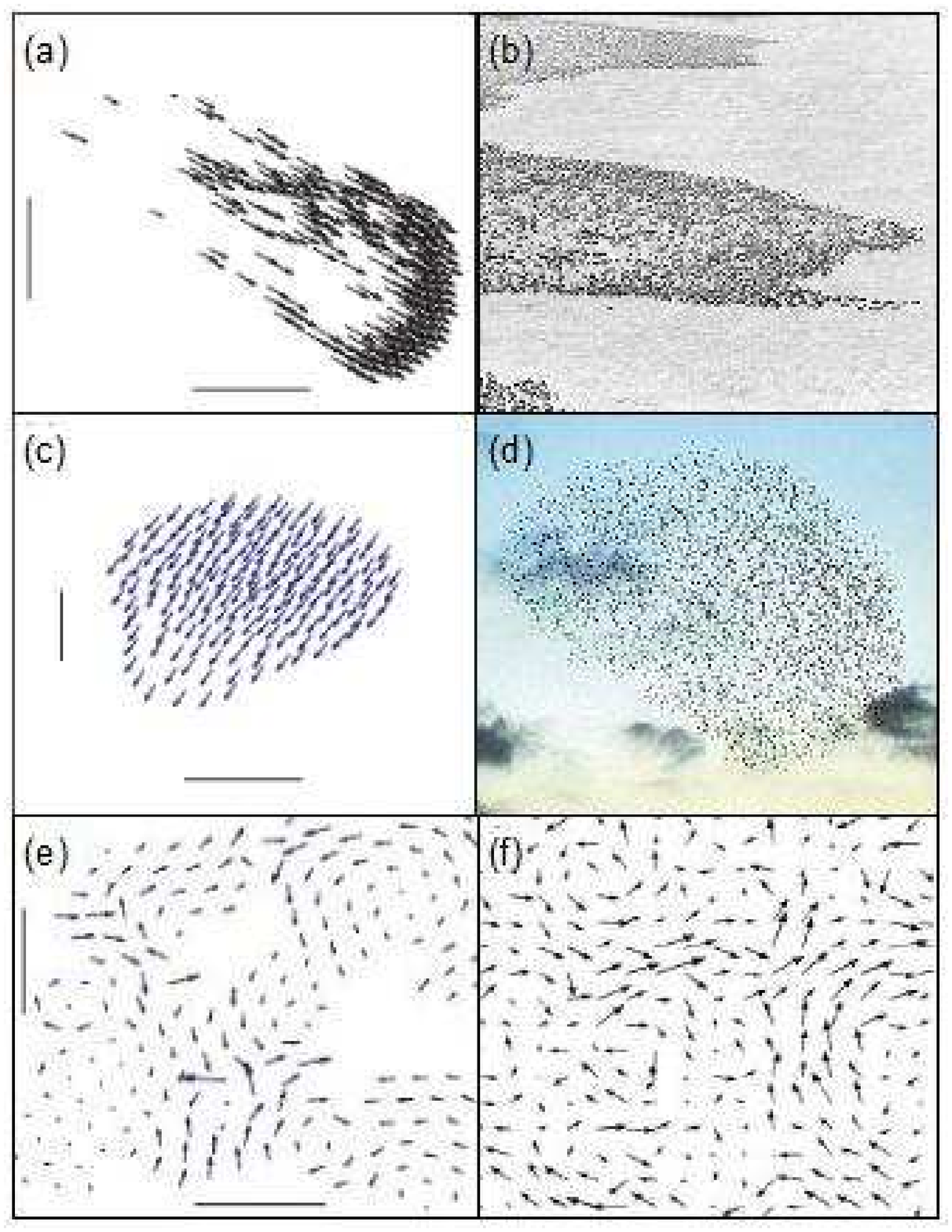}}
\caption{(a) Simulated stationary fluid state, with flow lines observed. (b) Experimental image of E.coli bacterium exhibiting fluid-like behaviour \cite{Ebac} , (c) Simulated single-flock MP state exhibiting distinctive teardrop structure. (d) A typical amorphous bird flock \cite{PhysToday}, (e) Simulated single-flock LJP state exhibiting distinctive wavefront structure (f) Migrating wildebeest exhibiting typical wavefront structure \cite{Couzin02}.}
\end{center}
\end{figure}
\\
This wavefront mode of flocking is observed in nature at the macroscopic level, with herds of wildebeest (and other ungulates) often forming distinct patterns qualitatively similar to those observed in our simulations during their migratory marches.  These wavefront flocks have been examined in some detail \cite{Couzin02}, but are primarily only understood qualitatively, with few quantitative measurements of their characteristics having been made. A popular theory for this shape of flock is that the region inside the "arc" of the wavefront becomes overgrazed, while the region outside has more resources for the flocks constituents to feed upon.  When combined with the natural tendency for flock members to follow one another, a breaking of the directional symmetry thus leads to the characteristic wavefront.  This characteristic shape is seen in Fig. 4(a), and the qualitative agreement with our model is seen by comparing this to Fig. 4(b).  However, our models do not include the effects of resource gathering (feeding), yet the characteristic shape is still observed for moving hard-core interactions.  This indicates that instead of being purely a result of outside forces, the wavefront shape exhibited in moving ungulates could instead be caused by the interactions between the flock members themselves. Due to the slow speed on ungulates relative to one another, they are unlikely to undergo accidental collisions, and thus we believe their behaviour is more accurately described the hard-core potential we have used.  Therefore, the question of whether the characteristic structure of ungulate herds is caused by interactions with the inhomogeneous resource pool, or interactions with one another is a very pertinent one in the light of these results.\\

For the MP, as density increases past the critical point,  the flock splits into two or more independent flocks along the line of motion with the flocks travelling in the same direction.  This characteristic splitting in clearly seen by comparing the regular teardrop shaped flock in Fig. 3(b) with the three separate flocks in Fig. 3(d) which occur at lower densities.  The state observed in Fig. 3(d) is subsequently called the "train".  As the density further decreases, there is a breaking of the velocity directional symmetry, resulting in multiple independent flocks travelling in different directions.  This holds true for all moving phases and noise regimes that were analyzed in our simulations.  Additionally, once the initial density is low enough for the directional symmetry to break, the independent flocks are free to move around and merge over time, with these mergers being observed in many of the simulations analyzed in the low density regime.\\

The single-flock MP phase is of particular interest, as behaviour similar to this is exhibited in many biological flocks.  In fact, spheroidal or amorphous flocks are amongst the most commonly encountered and include many types of bird flocks, fish schools and insect swarms.  The qualitative similarities between the modelled single flock MP phase and an experimental observed bird flock can be seen in figure 4(c)-4(d).  However there is a singificant lack of experimental data for flocking in 3D (fish and bird schools) and it is yet to be confirmed that similar spherical structures will form the 3D extension of the MP modelled used here.\\

It is also important to note the existence of vortex-like behaviour over a wide range of densities in our stationary state simulations $A=0$ using the MP.  This vortex behaviour warrants further analysis in the long-term due to its correspondence with observed behaviour in bacteria, in particular Dictyostelium \cite{Rappel}, Daphnia \cite{Daph} and E.Coli \cite{Ebac}.  However, the vortex structures are quite sensitive to changes in noise and are only observed in the low noise regimes.  In the medium and high noise cases, the vortex structure is broken down by thermal motions.  The fact that this state exists in our simulations at all is a good indication for the generality of the models presented here.  Good qualitative agreement is found between our simulated case and experimental observations in E.Coli\cite{Ebac} in particular.  This is shown in Fig. 4(e) and Fig. 4(f).\\

Fig. 5 and Fig. 6 shows a set of phase diagrams for both the LJ and Morse based models with $\eta$ plotted against $l=\frac{1}{\rho^{1/2}}$ in both cases for fixed values of $A=1$ (moving states).  Low noise corresponds to $\eta=1$ for both models.  In the MP case, medium noise corresponds to $\eta=20$ and high noise to $\eta=40$ and in the LJP case medium noise corresponds to $\eta=10$ and high noise to $\eta=20$.\\

As stochastic noise, $\eta$ increases, the liquid-like behaviour of the flock becomes dominant, and for small values of $\eta$ the solid-like behaviour is dominant.  The formation of single flock or multi-flock states is highly dependent on the initial density $\rho$, even in the absence of boundary conditions.  For high values of $\rho$ in the respective models, a single flock state will form, as $\rho$ is decreased, the flock transitions into an intermediate phase, and then at lower values of $\rho$, multiple independent flocks are observed.
Fig. 5 and  Fig. 6 show the results for the MP  and LJP respectively in the moving regime $(A=1)$.  In both cases, three distinct phases are observed as density decreases.  In the MP, as $\rho$ decreases the flock splits along the line of motion, resulting in a phase transition between the single flock and train phases.  Further decrease in $\rho$ eventually results in a further breaking of directional symmetry, whereby multiple
independent flocks form each with random direction.  This is true in low (Fig. 5(a)), medium (Fig. 5(b))and high (Fig. 5(c)) noise regimes, although the actual values of $\rho_{crit}$ vary slightly depending on the amount of stochastic noise in the system.  Figure 6 shows similar behaviour in the case of the LJP.  In this case however the transitionary state is quite different.  As shown in Fig. 3(d) as density decreases, the flock tears along the middle, resulting in splitting of the flocks wavefront.  As density decreases further these fragmented flocks cease to interact and the result in a multi-flock state with a number of independent flocks moving in free space, as for the MP case.  Additionally, in the case of the LJP, the position of the phase transitions described above is more difficult to resolve than for the MP, resulting in larger regions of uncertainty and making the nature of the phase transition less clear.  As for the LJP, the same general behaviour is observed in low (Fig. 6(a)), medium (Fig. 6(b))and high (Fig. 6(c)) noise regimes, despite the actual values for $\rho_{crit}$ varying slightly.
Thus, the positions of the phase transitions in the moving regime are not highly sensitive to noise, and in fact, variation of the stochastic noise seems to have only a small effect on the position of the transitions in phase space.\\
\begin{figure}
\begin{center}{\includegraphics[height=4.2cm, width=6.1cm]{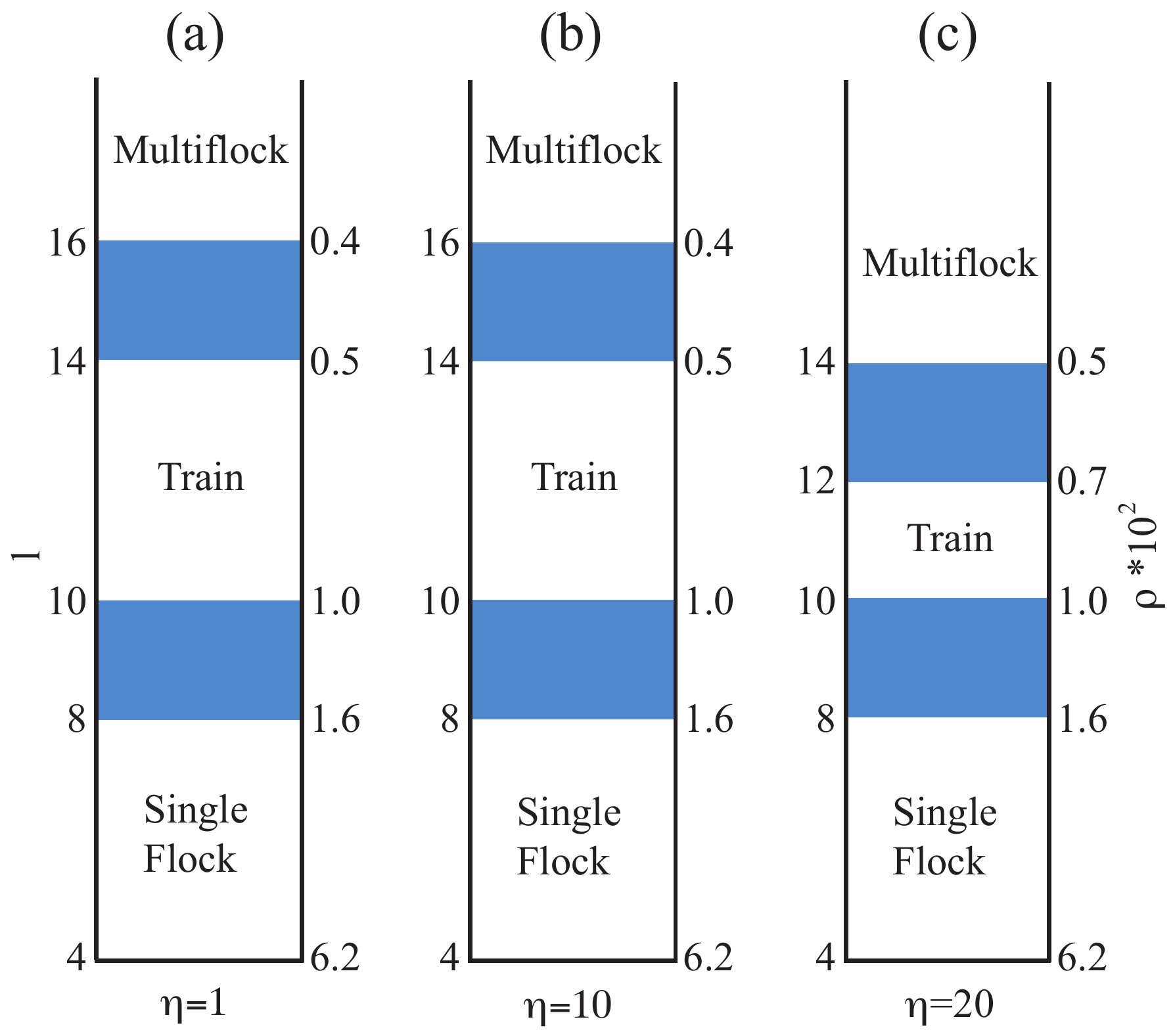}}
\caption{Phase Diagram for MP in the moving regime $(A=1)$ plotted against $l=\frac{1}{\rho^{(1/2)}}$ for three different cases corresponding to, (a) low $(\eta=1)$, (b) medium $(\eta=20)$ and (c) high $\eta=40$ values of stochastic noise.}
\end{center}
\end{figure}
\begin{figure}
\begin{center}{\includegraphics[height=4.2cm, width=6.1cm]{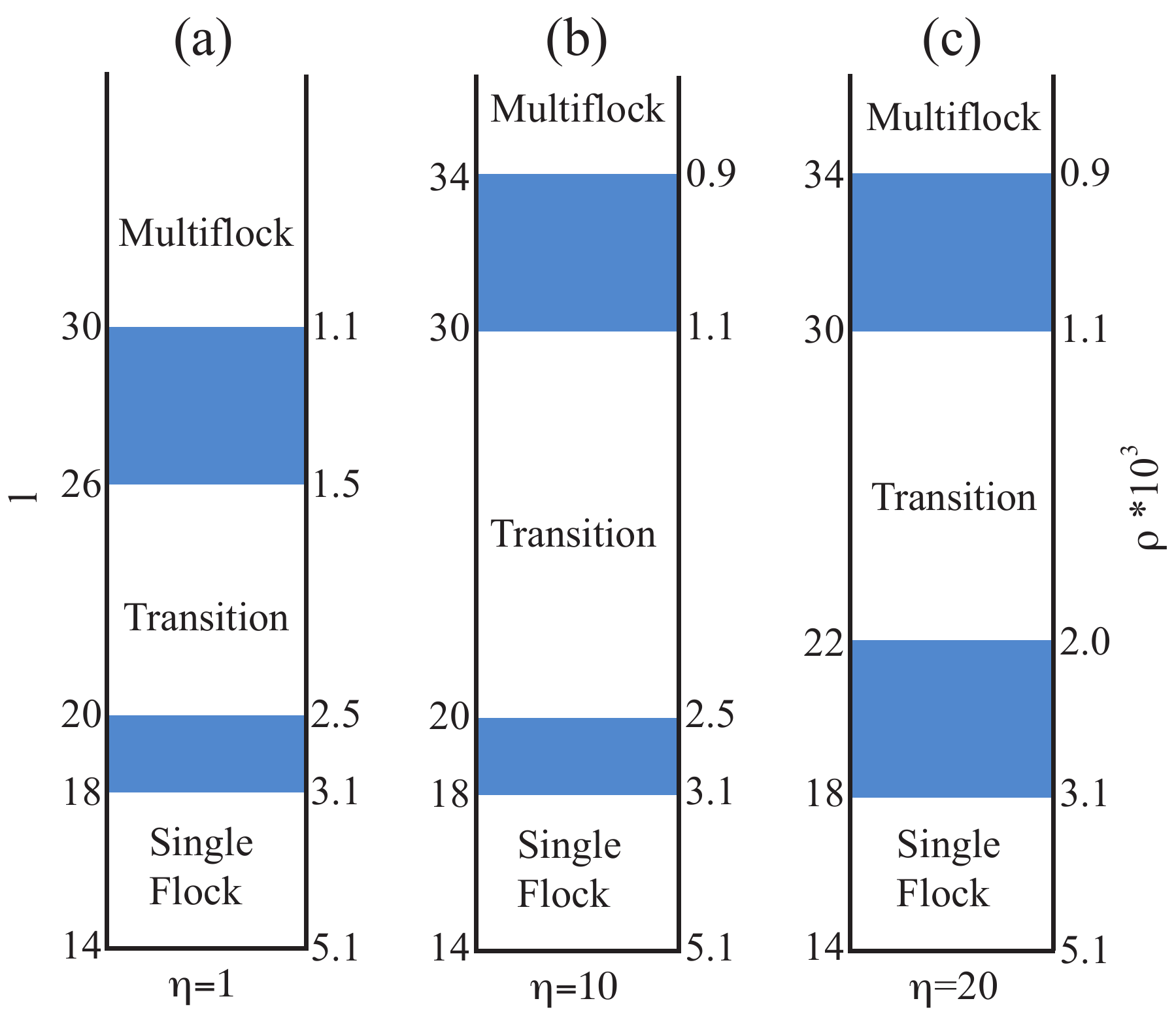}}
\caption{Phase Diagram for LJP in the moving regime $(A=1)$ plotted against $l=\frac{1}{\rho^{(1/2)}}$ for three different cases corresponding to, (a) low $(\eta=1)$, (b) medium $(\eta=10)$ and (c) high $(\eta=20)$ values of stochastic noise.}
\end{center}
\end{figure}
\\
  In the stationary regime where $A=0$ the phase transitions are a somewhat simpler.  In the MP case, two phases are observed with a discontinuous phase transition between the single and multi-flock state.  There is no intermediate phase between the single and multi-flock states, as we observed in the moving regime.  At high densities a single flock is formed, and as the density $\rho$ is decreased, there is little change in this single flock state until the critical value is reached.  There is some uncertainty as to the exact value at which the phase transition takes place, but Table 1 shows the approximate values of $l$ and $\rho$ where this takes place.  At values lower than $\rho_{crit}$ two independent flocks are observed.  Thereafter, further decreases in $\rho$ lead to an increase in the number of independent flocks present.  In this case it is important to note that position of the phase transition  is more sensitive to stochastic noise than in the moving states, with a consistent increase in the critical value of $\rho$ which results in a transition between the single and multiple flock states as noise, $\eta$ increases.  For the case of the LJP, a decrease in density simply leads to an increasingly dispersed flock, but does not result in the formation of multiple independent flocks.  It is not well understood why this should be the case, but it is believed that it is a consequence of the random placement algorithm used in our simulations in which boids are prevented from being placed too close together in the case of the LJP due to the infinite hard-core, since this has the effect of reducing the amount of initial inhomogeneities which subsequently form into the multiple independent flocks in the MP simulations.\\
  \begin{table}[!htb]
   \begin{center}
    \begin{tabular}{| c | c | c |}
    \hline
    Noise($\eta$)  & Side Length($l_{crit}$) &  Density($\rho_{crit}*10^{-3}$)\\ \hline
    1 & 20 - 22 & 2.5 - 2.1 \\ \hline
    20 & 18 - 20 & 3.1 - 2.5 \\ \hline
    40 & 16 - 18 & 3.9 - 3.1 \\
    \hline
    \end{tabular}
    \caption{Phase Transition points in the stationary flock regime $(A=0)$ for MP case at low, medium and high values of stochastic noise.}
   \end{center}
  \end{table}

\section{Conclusion}

It has been shown that both the Lennard-Jones and Morse potentials, when coupled with an alignment potential, are capable of replicating a wide range of flocking behaviour in free space, including those phases observed in Vicsek's original flocking paper, that have previously only been studied with the aid of periodic boundary conditions.  Additionally, it has been shown that while their are significant differences in the quantitative details, the bulk features of both the hard core LJP and the soft core MP are the quite similar in most key regards.  This indicates a kind of "potential independence" which should operate between all potentials which have the same key features of short-range repulsion, a preferred nearest-neighbour distance, and long range attraction that weakens with increasing inter-boid distance.\\

However, significant differences appear upon deeper examination of the two Potential types.  The behaviour around the phase transitions differs in that the transition between the single and multi-flocks states occurs discontinuously at some single value of $\rho$ whereas in the LJP, the breakdown of the single flock is gradual, taking place over a wide range of $\rho$ making it difficult to identify a single point where this phase transition occurs.\\

Additionally, the transition from single to multi-flock states happens in two stages.  In the first stage , the flock is observed to split along the line of motion, with all flocks continuing to travel in the same direction.  The second stage occurs as $\rho$ is further decreased, resulting in a breaking of directional symmetry, which leads to a number of independent flocks all travelling in effectively random directions.  Once this state is reached, the individual flocks interact only minimally until they come into close proximity with one another.  A number of behaviours can be observed when these independent flocks come into contact with one another including flock mergers, scattering of flocks off one another and flock collisions. All of these interactions have been observed in our model but have not yet fully investigated.\\

Finally, several vortex states were observed in the low noise regime of the stationary states for the MP.  In the LJP these vortices were not observed, but instead solid lattice behaviour was observed in the corresponding regimes for this class of potential.  The observation of these vortices is important since it is indicative of the ability of our model to effectively simulate bacterial behaviours which have already been observed in experimental situations and have been more extensively measured than most macroscopic examples of flocking.\\
Although a number of important observations have so far been made using these models, there is still much scope for future analysis of a number of flocking properties.  The models presented have proven very effective in qualitatively simulating the behaviour of a wide range of different flocking behaviours over the entire spectrum of length scales, which was the original aim when they were developed.\\

\end{document}